\documentclass[12pt]{article}

\usepackage{amsfonts}
\usepackage{bbm}

\newcommand{\bmat}{\left(\begin{array}}
\newcommand{\emat}{\end{array}\right)}
\newcommand{\uno}{\mathbbm{1}}
\def\NPB#1#2#3{Nucl. Phys. B{#1} (#2) #3}
\def\PLB#1#2#3{Phys. Lett. B{#1} (#2) #3}

\def\a{\alpha}
\def\ap{\alpha^{\prime}}

\def\g{\gamma}

\def\-{\hphantom{-}}
\def\ov{\overline}
\def\s2{\frac{1}{\sqrt2}}

\def\beq{\begin{equation}}
\def\eeq{\end{equation}}
\def\beqa{\begin{eqnarray}}
\def\eeqa{\end{eqnarray}}
\def\im{{\rm Im \,}}
\def\re{{\rm Re \,}}

\def\Tr{{\rm Tr \,}}

\def\T{{\rm T}}
\def\Z{{\mathbb Z}}

\def\cg{{\cal G}}
\def\cam{{\cal M}}
\def\cn{{\cal N}}

\def\nft{\widetilde{N}_{flux}}
\def\deq#1{\mbox{$D$=#1}}
\def\neq#1{\mbox{$\cn$=#1}}
\def\Dsl{\,\raise.15ex\hbox{/}\mkern-13.5mu D} 

\newcommand{\mathsmaller}[1]{\mbox{\footnotesize$#1$}}
\begin{document}
\pagestyle{empty}
\begin{flushright}
{\tt IFT-UAM/CSIC-04-58}
\end{flushright}
\vspace*{2cm}

\vspace{0.3cm}
\begin{center}
{\LARGE \bf $\Z_N$ Orientifolds with Flux }\\[1cm]
Anamar\'{\i}a Font
\footnote{On leave from Departamento de F\'{\i}sica, Facultad de Ciencias,
Universidad Central de Venezuela, A.P. 20513, Caracas 1020-A, Venezuela.}\\[0.2cm]
{\it  Instituto de F\'{\i}sica Te\'orica C-XVI,
Universidad Aut\'onoma de Madrid,\\[-1mm]
Cantoblanco, 28049 Madrid, Spain. }\\[2cm]
\normalsize{\bf Abstract} \\[8mm]
\end{center}

\begin{center}
\begin{minipage}[h]{15.5cm}
\normalsize{
We compute the flux induced tadpole and superpotential in
various type IIB $\Z_N$ compact orientifolds in order to 
study moduli stabilization. We find supersymmetric vacua with $g_s < 1$
and describe brane configurations with cancelled tadpoles.
In some cases moduli are only partially fixed unless anti D3-branes
are included.
}
\end{minipage}
\end{center}

\newpage

\setcounter{page}{1} \pagestyle{plain}
\renewcommand{\thefootnote}{\arabic{footnote}}
\setcounter{footnote}{0}

\section{Preamble}

A most attractive feature of compactifications with fluxes is the
existence of a moduli dependent potential with minima at finite
values. Flux induced moduli stabilization in type IIB orientifolds
on $\T^6$, ${\rm K3} \times \T^2$, $\T^6/\Z_2 \times \Z_2$, and
Calabi-Yau threefolds, has been largely discussed in recent times 
\cite{drs, kst, fp, tt, blt, cu, lrs1, ad, gktt, dd, gkt, mn, cq}.  
In this note we consider type IIB $\neq1$ orientifolds with internal space
$\T^6/\Z_N$, studied only briefly up to now \cite{cu}, which in fact
provide a simpler setup in which moduli stabilization can be
described in detail. Moreover, full models with tadpole cancellation
can be constructed. We will focus on the behavior of complex structute moduli
and the dilaton-axion $\tau= C_0 + i e^{-\phi}$. We have no new insights 
about fixing K\"ahler moduli so they will be dropped from the analysis.  

In Type IIB compactifications it is natural to turn on fluxes of the NS-NS and R-R 
3-forms. These fluxes must satisfy the quantization conditions
\beq
\frac1{(2\pi)^2 \ap} \int_\gamma \, F_3 \in \Z \quad ; \quad 
\frac1{(2\pi)^2 \ap} \int_\gamma \, H_3 \in \Z  \ ,
\label{fluxquant}
\eeq
for any 3-cycle $\gamma$ in the internal space $\cam$. It is
useful to introduce the combination $G_3= F_3 - \tau H_3$.
The equations of motion require that $G_3$ be imaginary self-dual (ISD), 
i.e. ${}^*G_3= i G_3$
\cite{gkp}. 

The fluxes induce a tadpole for the R-R 4-form with coefficient
\beq
N_{flux} = \frac1{(2\pi)^4 \a^{\primeÂ\, 2}} \int_{\cam} \, H_3 \wedge F_3 \ .
\label{tadflux}
\eeq
In $\Z_N$ toroidal orientifolds with
orientifold action including a reflection $I_6$ of all six internal
coordinates there are 64 O3-planes, and typically D3-branes, that 
contribute to the tadpole. We consider only O3-planes with neither NS-NS nor
R-R backgrounds, thus with charge $-1/2$ (in units such that a D3-brane has charge 1). 
This requires that in the flux quantization (\ref{fluxquant}) 
all integers be even \cite{kst, fp, cu}. The $C_4$ tadpole cancellation
condition is then
\beq
N_{{\rm D}_3} + \widetilde{N}_{flux} = 32 \ ,
\label{c4tad}
\eeq
where $N_{{\rm D}_3}$ is the net number of D3-branes. Here $\widetilde{N}_{flux}$  
is computed on the torus, this is appropriate because the number
of dynamical 3-branes is not $32/2N$. In general there are twisted tadpoles
whose cancellation requires a non-zero number of D3-branes pinned down at
the origin, i.e. the point fixed by the full orientifold group. For $N$ even there 
are further O7-planes and D7-branes and the $C_8$ tadpole must be cancelled.  

The fluxes also generate a superpotential \cite{gvw}
\beq
W = \int_{\cam} \, G_3 \wedge \Omega \ ,
\label{wflux}
\eeq
where $\Omega$ is the holomorphic (3,0) form. The superpotential depends on the
dilaton and the complex structure moduli denoted $U_\a$. The ISD condition on $G_3$
is equivalent to demanding \cite{gkp}
\beq
D_\tau W \equiv\partial_\tau W +  (\partial_\tau K) W=0 \quad ; \quad 
D_\a W \equiv\partial_\a W +  (\partial_\a K) W=0 \ ,
\label{isdcond}
\eeq
where
\beq
K=-\log[-i(\tau - \ov{\tau})] - \log\left(-i \int_{\cam} \ \Omega \wedge 
\ov{\Omega} \right)
\label{kpot}
\eeq 
is the tree-level K\"ahler potential for $\tau$ and $U_\a$. We will 
not include K\"ahler moduli in the analysis. However, we implicitly assume
that their K\"ahler potential is of no-scale form so that (\ref{isdcond})
corresponds to the minimum of the supergravity potential. Generically
$W$ does not vanish at the minimum, thus supersymmetry is broken by the
F-term of the K\"ahler moduli. This soft supersymmetry breaking is 
measured by $\exp Ã{\cg}$, where $\cg=K + \log |W|^2$.

As usual we define complex coordinates $z^a = x^a + i x^{a+3}$, $a=1,2,3$, 
on which the $\Z_N$ action is $\theta: z^a \to e^{2i\pi v_a} z^a$.
To preserve supersymmetry we specifically choose $\sum_a v_a =0$. The torus lattice is
denoted $\Lambda$, and its basis $e_i$, $i=1, \cdots, 6$. Since $\Lambda$
must be $\Z_N$ symmetric, its deformation parameters are restricted, being actually
in one-to-one correspondence with the untwisted K\"ahler and complex structure 
moduli \cite{cgm, ek}. In fact, in these orbifolds there is at most one untwisted 
$U$ modulus, allowed when $N$ is even and, say, $v_3=-1/2$. 

We use conventions in which 
\beq
\Omega= \lambda \, dz^1 \wedge dz^2 \wedge dz^3 \ .
\label{omegadef}
\eeq
Since the $x^i$ have flat metric $\delta_{ij}$, 
$\Omega$ is imaginary anti-self-dual, i.e.  ${}^*\Omega= -i \Omega$. 
The normalization factor $\lambda$ generically has units
of $({\rm length})^{-3}$ and will be chosen so that $K$ and $W$
do not depend on the K\"ahler moduli. This is just a matter of
convenience. The physically relevant quantity $\cg$
is independent of $\lambda$, and the K\"ahler moduli.
When $N$ is even, and $v_3=-1/2$, there is also an untwisted 
invariant (2,1) form, namely
\beq
\xi=  \lambda \, dz^1 \wedge dz^2 \wedge d\bar{z}^3 \ .
\label{xidef}
\eeq
Notice that $\xi$ is ISD. 

In general, $F_3$ and $H_3$ are linear combinations of all $b_3=2 + 2h_{12}$ 3-forms
but in this note we only turn on fluxes along untwisted 3-forms such as
$\Omega$, $\xi$, and their complex conjugates. Such fluxes are also primitive.
The K\"ahler form, $J= \sum_a J_a dz^a \wedge 
d\bar{z}^a$, satisfies $J\wedge \Omega=0$ and $J\wedge \xi=0$.

\section{Examples}
\label{ejem}

In this section we determine the flux induced quantities in various
$\T^6/\Z_N$ orientifolds. We will discuss to what extent the fluxes fix
the values of the moduli and comment on the effect in the open string spectrum.

We restrict to $\Z_N$ orbifolds in which there is no contribution to $h_{21}$ from 
twisted sectors. In this situation, the symplectic basis of 3-cycles does not 
include fractional ones that might require flux integers to be multiple of some 
minimal quantum \cite{blt}. We only need demand even integers in order to avoid exotic 
O3-planes as mentioned before. 

\subsection{$\T^6/\Z_3$}
\label{zz3}

The torus lattice has nine free parameters, all corresponding to untwisted
K\"ahler moduli. To simplify we choose $\Lambda$ to be the product of
three $SU(3)$ root lattices (up to size) orthogonal to each other. This means
$e_i \cdot e_{i+1} = -R_i^2/2$, $i=1,3,5$, and all other mixed $e_i \cdot e_j=0$.
A suitable normalization for $\Omega$ is $\lambda^{-1}=R_1 R_3 R_5$.
Since $h_{21}=0$, there are only two homology 3-cycles and two harmonic 3-forms,
namely $\Omega$ and $\ov{\Omega}$. Hence we can write 
\beq
F_3 = a_F \ov{\Omega} + c.c. \quad ; \quad 
H_3 = a_H \ov{\Omega} + c.c.  \ .
\label{fhexpz3}
\eeq 
To determine the coefficients we must impose the
quantization conditions (\ref{fluxquant}) and to this end we need a basis 
of 3-cycles.

As in \cite{bbkl} we define the toroidal 3-cycles
\beq
\pi_{ijk} = e_i \otimes e_j \otimes e_k \ ,
\label{picyc}
\eeq
where the $e_i$ are the 1-cycles corresponding to the lattice basis.
Invariant combinations are systematically found by taking $\Z_N$ orbits
of the $\pi_{ijk}$ \cite{bbkl}. Given the $\Z_3$ action, $\theta e_i = e_{i+1}$,
$\theta e_{i+1} =-e_i- e_{i+1}$, $i=1,3,5$, one readily checks that there are
just two invariant independent 3-cycles. A convenient basis is provided by the 
orbits of $\pi_{135}$ and $\pi_{136}$. Explicitly,
\beqa
\gamma_1 & = & -\pi_{136} -\pi_{145} -\pi_{146} - \pi_{235} -\pi_{236} - \pi_{245}
\nonumber \\[0.2cm]
\gamma_2 & = & \pi_{136}  + \pi_{145} + \pi_{235} + \pi_{135} - \pi_{246} \ .
\label{invcycz3}
\eeqa
Besides, $\gamma_1 \cap \gamma_2 =1$ so that the basis is symplectic. The periods
of $\Omega$, i.e. the integrals $Y_A= \int_{\gamma_A} \Omega$, are easily calculated
to be $Y_1=1$ and $Y_2=e^{2i\pi/3}$.
Using (\ref{fluxquant}) we then obtain
\beq
a_H = \frac{1}{2 \sqrt{3}}[\sqrt3 \ell_1  + i(2\ell_2 + \ell_1)] 
\quad ; \quad  \ell_1, \ell_2 \in \Z  \ .
\label{afz3}
\eeq
For $a_F$, just replace $\ell$ by $k$. To avoid cluttering we omit factors of 
$(2\pi)^2 \ap$ that can be reinserted in the end. 
An equivalent way to proceed is to expand
the fluxes in the dual cohomology basis $\Sigma_A$ such that 
$\int_{\gamma_B} \Sigma_A = \delta_A^B$. Then, $F_3= k^A \Sigma_A$ and 
$H_3= \ell^A \Sigma_A$. {}From (\ref{afz3}) one can easily deduce the $\Sigma_A$
in terms of $\Omega$ and $\ov{\Omega}$ and compute the intersection $\eta_{12}=-1$,
where $\eta_{AB}=\int_{\cam} \Sigma_A \wedge \Sigma_B$.

For the flux induced tadpole we find
\beq
\widetilde{N}_{flux}  =  3(k_1 \ell_2 - k_2 \ell_1) \ . 
\label{nfluxz3}
\eeq
Clearly, $\nft= 3\ell^A \eta_{AB} k^B$. The factor of 3 takes into acount
that $\nft$ is computed on the torus. The superpotential is 
$W=i \sqrt3 (a_H \tau - a_F)$.
Imposing the ISD condition as $D_\tau W=0$, or as cancellation 
of the coefficient of $\Omega$ in $G_3$, gives 
\beq
\tau = \frac{ 2k_2 + k_1 + i\sqrt3 k_1 } { 2\ell_2+\ell_1 + i \sqrt3 \ell_1 } 
\ .
\label{tauz3}
\eeq
Notice that requiring ${\rm Im \tau} > 0$ implies $\nft > 0$.  
Recall that the flux integers are even, thus $\nft$ is quantized in multiples of 12. 

Another constraint on the fluxes is the validity of the perturbative expansion. Using 
(\ref{nfluxz3}) and (\ref{tauz3}) we can write the string coupling $g_s=e^{\phi}$ as
\beq
g_s = \frac{2\sqrt3 (\ell_2^2 + \ell_1^2 + \ell_2 \ell_1)}{\nft} \ .
\label{gsz3}
\eeq
This immediately shows that there are no solutions with $\nft=12$ and $g_s < 1$.
Furthermore, there is basically just one solution with  $\nft=24$ and $g_s < 1$. º 
For instance, using a vector notation, the fluxes 
$\ell=(0,2)$, $k=(4,-2)$, fix the dilaton at $\tau_0=i\sqrt3$, and 
at the minimum $|W_0|=4\sqrt3$. All other choices 
are the same up to $SL(2, \Z)$ transformations.
With larger $\nft$ the number of solutions with $g_s <1$ obviously increases.
For instance, for $\nft=48$ there are vacua with $g_s=\sqrt3/6$ and $g_s=\sqrt3/2$.
However, according to (\ref{c4tad}), anti D3-branes must be added to absorb the extra charge, 
thus breaking supersymmetry in an explicit way.

In the $\Z_3$ orientifold with $\Omega^\prime= \Omega (-1)^{F_L} I_6$,
twisted tadpole cancellation requires that at least 8 D3-branes be placed at the origin $O$.
This can be seen directly from the condition $\Tr \g_{\theta,3,O}=-4$, where
$\g_{\theta,3,O}$ is the embedding of the orbifold action on the Chan-Paton labels.
In turn this condition can be obtained applying T-duality to the regular $\Omega$
orientifold with D9-branes in which $\Tr \g_{\theta,9}=-4$ \cite{abpss, afiv}. The smallest
such $\g_{\theta}$,  with $\g_{\theta}^3=\uno$, is of dimension 8.

An alternative view is the following. In the regular $\Omega$ orientifold the D9 gauge
group is $SO(8) \times U(12)$, which can be broken by Wilson lines \cite{afiv}. In particular,
there is a family of discrete Wilson lines that gives $SO(8-2n) \times U(12-2n) \times U(n)^3$,
$n=0, \cdots, 4$. In the T-dual picture the first two factors arise from $(32-6n)$ D3-branes
staying at the origin while the last factor is due to $3n$ branes located at some other fixed
point $P$ of $\theta$, for which $\Tr \g_{\theta,3,P}=0$ (the remaining $3n$ branes are at the
image of $P$ under $I_6$). Since all branes are placed at $\Z_3$ fixed points the rank
of the group is not reduced. Observe that we can remove at most 24 D3-branes from $O$. These could
be arranged as just described or they could be sent fully to the bulk as 4 dynamical 3-branes
that could be traded by flux.

The upshot is that with $\nft=24$ all tadpoles can be cancelled by placing 8 
D3-branes at the origin. They have gauge group $U(4)$ and three chiral multiplets in 
the ${\bf 6}$. This chiral content is clearly anomaly free. 

With lower $\nft$ more interesting brane configurations can be designed. A 3-family
$SU(5)$ model with $\nft=12$ and 20 D3-branes is described in \cite{cu}. 

\subsection{$\T^6/\Z_7$}
\label{zz7}
                                                                                                  
The torus lattice has three K\"ahler moduli encoded in the $R_i$ entering
in the lattice vectors \cite{cgm}. As $\Omega$ normalization we 
take $\lambda^{-1}=R_1 R_3 R_5$.
In this case $h_{21}=0$ so the fluxes are of the form (\ref{fhexpz3}).
Using $\theta e_i = e_{i+1}$, $i=1,\cdots,5$, $\theta e_6 = -e_1-e_2-e_3-e_4-e_5-e_6$,
shows that there are just two invariant independent 3-cycles. As basis we
take $\gamma_1, \gamma_2$ equal to the orbits of $\pi_{123}$ and $\pi_{124}$
respectively. These fulfill $\gamma_1 \cap \gamma_2 = 1$. The periods of $\Omega$
are $Y_1= -i \sqrt7$ and  $Y_2=(7 + i\sqrt7)/2$. It then follows
\beqa
a_H & = & \frac{1}{14}(2\ell_2 + \ell_1 - i\sqrt7 \ell_1) \ ,
\nonumber \\[0.2mm]
\nft & = & 7 (k_1 \ell_2 - k_2 \ell_1) \ .
\label{allz7}
\eeqa
The superpotential is  $W=i 7\sqrt7 (a_H \tau - a_F)$. 
The ISD condition leads to
\beq
\tau = \frac{ 2k_2 + k_1 + i\sqrt7 k_1 } { 2\ell_2+\ell_1 + i\sqrt7 \ell_1 } \ .
\label{tauz7}
\eeq
As before, ${\rm Im \tau} > 0$ implies $\nft > 0$.
Without anti D3-branes the maximum $\nft$ is 28. There is just one vacuum with $g_s < 1$,
achieved choosing, say $\ell=(0,2)$, $k=(2,0)$. Then, $\tau_0=(1+i\sqrt7)/2$, $|W_0|=14$.

According to (\ref{c4tad}), to cancel the untwisted tadpole, 4 D3-branes are needed. 
Twisted tadpole cancellation requiring $\Tr \g_{\theta,3,O}=4$ is then satisfied 
placing the branes at the origin. They give rise to group $SO(4)$ without matter.
In the regular orientifold, the D9 gauge group $SO(8) \times U(4)^3$ can be broken by 
a discrete Wilson line to $SO(4) \times U(2)^7$. In the T-dual picture the $SO(4)$ comes 
{}from the 4 branes at the origin and $U(2)^7$ from $14$ branes located at another $\theta$ fixed
point $P$, for which $\Tr \g_{\theta,3,P}=0$ (the remaining $14$ branes being at the
orientifold image). The 28 D3-branes liberated from the origin can move completely
to the bulk and turn into flux.

The $SO(4)$ group of the D3-branes at the origin is pure \neq1 super Yang-Mills. 
This raises the possibility of adding a non-perturbative superpotential 
$W_{np} \sim \exp(8\pi^2 i\tau/{\mathsmaller {C(G)}})$ generated 
by gaugino condensation. However, for $SO(4)$ with Casimir ${\mathsmaller {C(G)}}=2$, 
the flux superpotential linear in $\tau$ dominates and the solution of 
$D_\tau (W + W_{np})=0$ stays at $\tau_0$. For gaugino condensation to produce 
a bigger effect one needs groups with a larger Casimir.

\subsection{$\T^6/\Z_6^\prime$}
\label{zz6}

The $\Z_6^\prime$ action has $v_1=\frac16$, $v_2=\frac13$, $v_3=-\frac12$. It can be 
realized in four different lattices \cite{ek}. Here we work with $\Lambda$ 
of $SU(6) \times SU(2)$.  All possible lattices allow five real deformation
parameters, two of them corresponding to one untwisted complex structure modulus.
The number of twisted moduli depends however on the lattice. In our case
there are no twisted 3-forms, i.e. $h_{12}=1$ \cite{ek}. 

The lattice vectors satisfy $\theta e_i= e_{i+1}$, $i=1,\cdots, 4$, 
$\theta e_5=-e_1-e_2-e_3-e_4-e_5$. Following \cite{cgm} we write them as
\begin{center}
\footnotesize
\begin{tabular}{lclclcl}
$e_1$ &=& $(R_1, R_3, R_5,0,0,0)$ & \ ; \ & $e_4$ &=& $(-R_1, R_3, -R_5,0,0,0)$ \\
$e_2$ &=& $\frac12(R_1, -R_3, -2R_5,\sqrt3 R_1, \sqrt3 R_3,0)$ & \ ; \ & $e_5$ &=& 
$\frac12(-R_1, -R_3, 2R_5,-\sqrt3 R_1, \sqrt3 R_3,0)$  \\
$e_3$ &=& $\frac12(-R_1, -R_3, 2R_5,\sqrt3 R_1, -\sqrt3 R_3,0)$ & \  ; \  & $e_6$ &=&
$R_6(0,0,\cos\beta, 0,0,\sin\beta)$ \\
\end{tabular}
\end{center}
The untwisted complex modulus is simply
\beq
U= \frac{R_6}{\sqrt3 R_5} e^{i\beta} \ .
\label{umod}
\eeq
Again, $\lambda^{-1}=R_1 R_3 R_5$. 

The most general 3-form background is of type
\beq
F_3 = a_F \ov{\Omega} + b_F \xi + c.c. \quad ; \quad
H_3 = a_H \ov{\Omega} + b_H \xi + c.c.  \ .
\label{fhexpz6}
\eeq
It is easy to show that there are four toroidal invariant 3-cycles.
As basis we take $\gamma_A$, $A=1,\cdots,4$, to be the orbits of $\pi_{123}$, $\pi_{124}$, 
$\pi_{126}$ and $\pi_{136}$ respectively. The intersection matrix is unimodular.
The periods of $\Omega$ are $Y_1=-i\sqrt3$, $Y_2=3$, $Y_3=-\sqrt3 U$
and $Y_4=-3i U$. To obtain the periods of $\xi$ just replace $U$ by $\ov{U}$.
Imposing flux quantization leads to
\beqa
a_H & = & \frac{1}{12 {\rm Im \,} U} \left[ \ell_4 - i\sqrt3 \ell_3 - 
U(\sqrt3 \ell_1 + i \ell_2) \right] 
\nonumber \\[0.2cm]
b_H & = & \frac{1}{12 {\rm Im \,} U} \left[ -\ell_4 - i\sqrt3 \ell_3 +
U(\sqrt3 \ell_1 - i \ell_2) \right]
\label{abz6p}
\eeqa
The superpotential is $W=12i \sqrt3 {{\rm Im \,} U} (a_H \tau - a_F)$.
The tadpole coefficient is 
\beq
\nft = 6(k_1 \ell_4 - k_2 \ell_3 + k_3 \ell_2 - k_4 \ell_1) \ .
\label{nfluxz6p}
\eeq
Indeed, the dual 3-forms have intersections $\eta_{14}=-1$, $\eta_{23}=1$ and
$\eta_{AB}=0$ otherwise. Note that $\nft$ is quantized in multiples of 24.

The ISD condition reduces to
\beqa
&{}& \sqrt3 k_3 + k_2 \ov{U} - \sqrt3 \ell_3 \tau - \ell_2 \tau \ov{U} = 0 
\nonumber \\[0.2cm]
&{}& k_4 - \sqrt3 k_1 \ov{U} - \ell_4 \tau + \sqrt3  \ell_1 \tau \ov{U} = 0
\label{isdz6p}
\eeqa
We assume that $\im \tau \not= 0$ and $\im U \not= 0$, i.e. we exclude
infinite string coupling and degenerate torus. Then, equations (\ref{isdz6p})
can generically be solved to obtain the ratio $\im \tau / \im U$,
as well as the values of $\re \tau$ and $\re U$. There is a remaining equation
quadratic in $\im U$ that, depending on the fluxes, might not have real solutions.
In fact, for $\nft=24$ we find no solutions with $\tau$ and $U$ completely fixed.
The best alternative is to obtain a relation between them. Clearly, this can be 
easily achieved with fluxes such that one of the equations in (\ref{isdz6p}) is
trivial. For example, with $k=(2,0,0,0)$, $\ell=(0,0,0,2)$, $\tau=-\sqrt3 \,  \ov{U}$. 
Or with $k=(0,0,2,0)$, $\ell=(0,2,0,0)$, $\tau\ov{U}= \sqrt3$. 

With $\nft=24$, eq. (\ref{c4tad}) is satisfied by taking $N_{{\rm D}_3}=8$. 
In fact, one needs 8 D3-branes at the origin to cancel twisted tadpoles. This 
cancellation condition, as well as the open string spectrum, can be deduced adapting the
results in the $\Omega$ orientifold with D9 and D5-branes \cite{afiv}. The D3-branes
at the origin have gauge group $U(4)$ with two hypermultiplets in the ${\bf 6}$. There
are 32 D7-branes that can be located at the origin thus giving group 
$U(4)\times U(4)\times U(8)$ with charged hypermultiplets as in the 55 sector \cite{afiv}. 
The 37 matter consists of one hypermultiplet in
$(\ov{\bf 4};{\bf 4},{\bf 1}, {\bf 1}) + ({\bf 4};{\bf 1}, \ov{\bf 4}, {\bf 1})$.
The full matter content is anomaly free. 

With larger $\nft$ one does find vacua with $\tau$ and $U$ fixed. 
For example, for $\nft=48$ there is the family $k=(2,0,2,2n)$,  $\ell=(0,2,0,2)$, 
$|n| \leq 3$, with $\tau=(n+i\sqrt{12-n^2})/2$, $U=\tau/\sqrt3$. Notice that
$g_s < 1$ for $|n| \leq 2$. For  $|n|=3$, $g_s > 1$, and $\tau$ is fixed under 
S-duality. For  $|n| \geq 4$ these fluxes could only give a solution with
$\im \tau =\im U = 0$.  

\section{Final Comments}

The purpose of this note was to work out further examples of \deq4, \neq1, toroidal
orientifolds with fluxes. As we have seen, a rather explicit analysis of moduli
stabilization can be carried out. Moreover, D3/D7-brane configurations with
cancelled tadpoles, and the corresponding chiral spectrum, can be found. These
orientifolds can serve as starting points to build semi-realistic models and
study other flux effects such as soft supersymmetry breaking 
\cite{ciu1, Grana, lrs1, ciu2, jl, lrs2}.

In toroidal orientifolds the fluxes are limited to small values unless anti 
D3-branes are included to absorb the extra charge. These anti-branes break 
supersymmetry explicitly, in particular the spectrum is no longer supersymmetric
\cite{au}. On the other hand, $\ov{\rm D3}$-branes could have useful applications
as in the proposal to produce de Sitter vacua \cite{kklt}.
Note however that in our case the number of $\ov{\rm D3}$-branes needed to increase 
the fluxes appreciably is large, so they are not a small perturbation above
a supersymmetric vacuum. Moreover, there are flux-brane anhilation processes
that could not be sufficiently suppressed \cite{kpv}. 

Another possibility to increase the allowed values of $N_{flux}$ would be 
to add magnetized D9-branes that can balance the excess charge without breaking 
supersymmetry as observed recently \cite{ms}. Indeed, magnetized D9-branes
have been used to construct semi-realistic  models in the $\T^6/\Z_2 \times \Z_2$
orientifold with fluxes \cite{blt, cu, lrs1, ms, Cvetic, lrs2}.

The fluxed orientifolds in this note furnish neat examples in which a counting 
of vacua could be performed thoroughly. We have concentrated on
small fluxes and $g_s < 1$, and found that in this situation there are only 
very few vacua. In the $\Z_3$ and $\Z_7$ there are no complex structure moduli and the
values of $\tau$ are indeed determined as in the toy model of a rigid Calabi-Yau 
discussed in \cite{ad, dd}. The $\Z_6^\prime$ is the analog of a Calabi-Yau orientifold
with one modulus. For large fluxes the number of solutions of (\ref{isdz6p}) is
expected to agree with general predictions \cite{ad}. In the regime of small fluxes
one can just analyze (\ref{isdz6p}) directly.

Finally, for the $\Z_3$ and $\Z_7$ type IIB orientifolds a heterotic dual without fluxes
is known \cite{abpss,kks}. It would be interesting to track down the effect of fluxes in 
the heterotic side.


\vspace*{2cm}

\noindent
{\bf Acknowledgments : } 
I thank L. Ib\'a\~nez, F. Quevedo, S. Theisen and A. Uranga for useful
remarks.

\end{document}